\documentclass[runningheads, envcountsame, a4paper]{llncs}
\usepackage{graphicx}

\usepackage[misc]{ifsym}
\usepackage{booktabs}  
\usepackage{tabularx}  
\usepackage{multirow}  

\begin{document}
	\title{Match4Rec: A Novel Recommendation Algorithm Based on Bidirectional Encoder Representation with the Matching Task}
	\titlerunning{Match4Rec: A Recommendation Algorithm with Matching Task}
	%
	\toctitle{Match4Rec: A Novel Recommendation Algorithm Based on Bidirectional Encoder Representation with the Matching Task}
	
	\author{Lingxiao Zhang\inst{1}(\Letter) \and
		Jiangpeng Yan\inst{1} \and
		Yujiu Yang\inst{1} \and
		Li Xiu\inst{1}(\Letter)}
	\authorrunning{L. Zhang et al.}
	%
	\tocauthor{Lingxiao~Zhang, Jiangpeng~Yan, Yujiu~Yang, and Li~Xiu}
	
	\institute{Tsinghua Shenzhen International Graduate School, Shenzhen 518055, China \\
		\email{zhang-lx18@mails.tsinghua.edu.cn, li.xiu@sz.tsinghua.edu.cn}}
	\maketitle              
	\begin{abstract}
		Characterizing users' interests accurately plays a significant role in an effective recommender system. The sequential recommender system can learn powerful hidden representations of users from successive user-item interactions and dynamic users' preferences. To analyze such sequential data, the use of self-attention mechanisms and bidirectional neural networks have gained much attention recently. However, there exists a common limitation in previous works that they only model the user's main purposes in the behavioral sequences separately and locally, lacking the global representation of the user's whole sequential behavior. To address this limitation, we propose a novel bidirectional sequential recommendation algorithm that integrates the user's local purposes with the global preference by additive supervision of the matching task. Particularly, we combine the mask task with the matching task in the training process of the bidirectional encoder. A new sample production method is also introduced to alleviate the effect of mask noise. Our proposed model can not only learn bidirectional semantics from users' behavioral sequences but also explicitly produces user representations to capture user's global preference. Extensive empirical studies demonstrate our approach considerably outperforms various baseline models.  
		
	\end{abstract}
	%
	%
	%
	\section{Introduction}
	Recommender Systems can help users obtain a more customized and personalized recommendation experience by characterizing users exhaustively and mine their interests precisely. A widely used approach to building quality recommender systems in real applications is collaborative filtering (CF) \cite{ref_1}. But such a method takes users' shopping behaviors as isolated manners, while these behaviors usually happen successively in a sequence. Recently, sequential recommendations based on users' historical interactions have attracted increasing attention. They model the sequential dependencies over the user-item interactions (e.g., like or purchase) in sequences to capture user interests \cite{ref_2}. Two basic paradigms of the pattern have proliferated: unidirectional (left-to-right) and bidirectional sequential model. The former, including Markov Chains (MC) \cite{ref_3}, Recurrent Neural Networks (RNNs) \cite{ref_6} and self-attentive sequential model \cite{ref_8}, is more close to the order of interactions between users and items in many real-world applications, yet it is not sufficient to learn optimal representations for user behavior sequences. The latter, like BERT4Rec \cite{ref_9}, premeditates various unobservable external factors and does not follow a rigid order assumption, which is beneficial to incorporate context from both sides for sequence representation learning. However, the aforementioned bidirectional sequential model only relays on capturing the user's main purposes, which are reflected by relatively important items distributed in different local areas of the whole sequence. Therefore, there exited a limitation that these models cannot always conjecture the user's main purposes without the global knowledge, especially, when the sequence is quite short or the user just clicks something aimlessly.
	
	In this paper, we consider the user's entire sequential behavior as the supplement of the local purposes. We integrate the mask task with the matching task by the novel mask setting of the unambiguous user sequential representation during the training processing of the bidirectional model. The matching task which usually directly build the mapping between the user's whole behavior sequence and the targeted items, treats the recommendation problems as the matching problem and can measure the user's global preference on items \cite{ref_10}. The mask task \cite{ref_9,ref_11} is adopted to substitute the objective in unidirectional models for the bidirectional models. Some items in the users' behavioral sequences are masked in certain probability (e.g., replace them with a special token [mask]). Then, the recommender model predicts the ids of those masked items based on their surrounding context, which is a mixture of both the left and right context. To integrate the matching task in such a mask task, we use a special token ``[UID]'' to explicitly represent individual users, inspired by doc2vec \cite{ref_12}. Then we concatenate the user token with several item tokens from a sequence to train a bidirectional encoder model. Thus, our model can determine whether or not each users' semantic vector (i.e., the output of the user token) and items' semantic vectors (including positive samples and negative samples) are well matched. Because the output of the user token has merged various correlations among items in each sequence, this method can also be applied to variable-length pieces of sequences and expressly form the user representation. To alleviate the effect of mask noise in the training, we produce instances that only compute the loss function of the matching task between the original user behavioral sequence and items. Extensive experiments on four datasets show that our model outperforms various state-of-the-art sequential models consistently.
	
	In conclusion, the contributions of this paper are listed as follows: a) We integrate the matching task with the bidirectional recommender model by the novel mask setting during the training. In this way, user's local purposes and the global preference in the behavioral sequence can be combined to boost the performance of the recommender system. b) We propose a novel sample production method to alleviate the effect of mask noise in the training for matching task. c) Extensive experiments show that our model outperforms state-of-the-art methods on four benchmark datasets.
	
	\section{Related Works}
	In this section, we briefly introduce several works closely related to ours. We first discuss general recommendation, followed by sequential recommendation and the process of the matching task in the recommendation.
	
	As mentioned above, Collaborative Filtering (CF) \cite{ref_1} is one of the most widely used general recommendations that takes users' shopping behaviors as isolated manners. Recently, deep learning techniques have been introduced for general recommendation. Some researchers tried to use more auxiliary information (e.g., text \cite{ref_13}, images \cite{ref_14}, acoustic \cite{ref_15}) into recommendation systems. Some works focused on replacing conventional matrix factorization (NCF \cite{ref_16}) with neural networks(e.g., AutoRec \cite{ref_17} and CDAE \cite{ref_18}).
	
	Different from the above methods, sequential recommendation systems consider orders in users' behaviors. Early works adopted the Markov chain to model the transition matrices over user-item interactions in a sequence \cite{ref_3}. Then recurrent neural networks (RNN) are widely used for sequential recommendation \cite{ref_6}. Apart from RNNs, other deep learning models are also applied in sequential recommendation systems. For example, Caser \cite{ref_19} learns sequential patterns through both horizontal and vertical convolutional filters. Recently, the use of attention mechanisms in recommendation has got the substantial performance. SASRec \cite{ref_8} applies a two-layer Transformer decoder to capture the user's sequential behaviors in left-to-right order (i.e., Transformer language model). BERT4Rec \cite{ref_9} uses a two-layer Transformer decoder with the help of the Cloze task to achieve bidirectional information mining, which is closely related to our work. 
	
	Matching tasks in the recommendation is used to capture the user's global preference on items. The fundamental problem of matching tasks is the semantic gap because users and items are heterogeneous objects, and there may not be any overlap between the features \cite{ref_10}. To address the problem, matching tasks usually are performed at the semantic level. Thus, the strong representation ability of the models is the key to improving recommendation performance. Deep learning methods are widely used in the matching task because of their great potentials of abstracting representations for data objects \cite{ref_16,ref_17,ref_18}. In this paper, we use Transformer to unambiguously represent individual users and perform the matching task, aiming to model user's global preference in the sequence to get better recommendation performance.

	\section{Methodology}
	In this section, we introduce our model architecture and several detailed modules. Firstly, some important variables are defined as the following. Considering a user's interaction sequence $S^{u_i}=[v_1^{u_i},v_2^{u_i},v_3^{u_i},...,v_{|V|}^{u_i}]$, the next item $v_{n+1}^{u_i}$ needs to be predicted by the sequential recommendation algorithm, where $v_i^u\in V$, item set $V=\{v_1,v_2,v_3,...,v_{|V|}\}$, and $u_i\in U$, item set $U=\{u_1,u_2,u_3,...,u_{|U|}\}$. Predicted probability can be formalized as ${\rm p}(v_{n+1}^{u_i}=v|S^{u_i})$.
	
	\subsection{Model Architecture}
\begin{figure}
		\centering
		\includegraphics[width=0.65\textwidth]{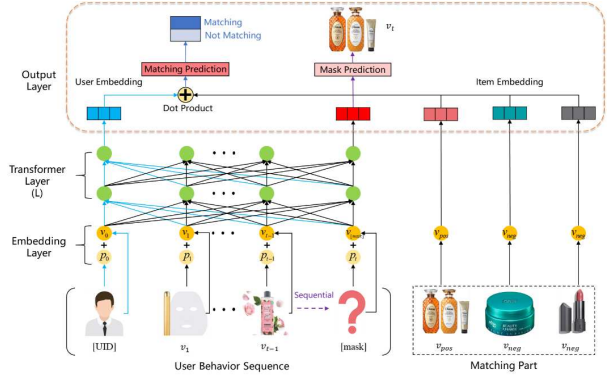}
		\caption{Our proposed model architecture.} \label{fig1}
	\end{figure}

	Our model architecture is shown in Fig. {\ref{fig1}}, which is made up of the embedding layer, transformer layers, and the output layer.
	
	In the embedding layer, the input sequential items are mapped into item embedding and position embedding. Note that to expressly build the user representation, we add a special token ``[UID]'' at the beginning of a sequence and share the weights from the item embedding with positive and negative items that are used in following matching task. After the embedding layer, we stack L Transformer layers to catch dependencies of items in each sequence. Different from other sequential models such as RNN, the self-attention mechanism directly computes dependencies of tokens in sequences rather than through accumulative dependencies in the last time. In the output layer, the model needs to predict masked items and determine if user-item pairs are well matched.
	
	The embedding layer and the output layer is specially designed for our proposed model, the transformer layers share the same structure of BERT \cite{ref_11}, which is used to process neural language sequence, and is also used in BERT4Rec \cite{ref_9} for recommendation problem.
	
	\subsection{Embedding Layer}
\begin{figure}
		\includegraphics[width=0.6\textwidth]{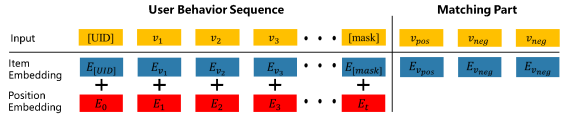}
		\centering
		\caption{Input representation of our model. The input embedding has two types: user embedding and item embedding.} \label{fig2}
	\end{figure}

	In our model, given a user's interaction sequence $S^{u_i}$, we set a restriction on the maximum sequence length $N$ to make sure our model can handle. In other words, we only consider the most recent $N$-1 actions (except special token ``[UID]'' at first of the sequence). The input embedding has two types: user embedding and item embedding. User embedding $E\in \textbf{R}^{N\times d}$ is made up by summing item embedding $E_v\in \textbf{R}^{N\times d}$ and position embedding $E_n\in \textbf{R}^{N\times d}$:
	\begin{equation}
	E = E_v+E_p.
	\label{eq1}
	\end{equation}
	Additionally, we use shared weights from item embedding $E_v$ to map one positive item (i.e., the last item) and $n$ random sampled negative items (non-interaction items) to item space. A visualization of this construction can be seen in Fig.~\ref{fig2}.
	
	Take the case shown in Fig. \ref{fig1} to help readers to understand. The consumer finally buys a shampoo, while we build a triple [shampoo, cream, lipstick] for the matching task. The shampoo is positive, the latter two items are negative.
	
	\subsection{Transformer Layer}
	As mentioned above, Transformer Layer is first proposed in \cite{ref_11} to build the bidirectional semantic representation for language understanding. Every Transformer layer is mainly constructed by a multi-head self-attention sub-layer and a feed-forward network \cite{ref_20}. Based on \cite{ref_20}, we employ a residual connection, layer normalization and dropout around each of the two sublayers to avoid overfitting model and vanishing gradient. The process is formulated as follow:
	\begin{equation}
	\textsl{g}(x)=x+\rm{Dropout}(\textsl{g}(\rm{LayerNorm}(x))),
	\label{eq2}
	\end{equation}
	where $\textsl{g} (x)$ represents the self-attention layer or the feed-forward network. 
	
	\subsection{Output Layer}
	In the output layer, we need to deal with two tasks: masked items prediction and matching prediction. For masked items, we get the final output $H_t^L$ after L Transformer layers, where $t$ means the masked item $v_t$ is at time step $t$. Softmax is employed as the activation function. The process is formulated as follow:
	\begin{equation}
	P_{Mask}(v_t)={\rm Softmax}(H_t^{L}W_p+b_p),
	\label{eq3}
	\end{equation}
	where $b_p$ is a learnable projection bias, $W_p$ is the projection matrix. In order to alleviate overfitting and reducing the model size, we make $W_p$ share weights from the item embedding matrix in the embedding layer. 
	
	For matching prediction, we extract the first final output $H_1^L$ (i.e., the final output of ``[UID]''), and a positive item $E_v^{pos}$ and n negative items $E_v^{neg}$ that have been mapped into item semantic space in the embedding layer. We calculate matching scores of a positive one and negative ones. The process is defined by:
	\begin{equation}
	Score_{pos}=E_v^{pos}\bullet H_1^L,
	\label{eq4}
	\end{equation}
	\begin{equation}
	Score_{neg}=(E_v^{neg}\bullet H_1^L)/n.
	\label{eq5}
	\end{equation}
	Note that the negative score is divided by the number of negative sampling $n$ to balance positive and negative score weights.
	
	\subsection{Model Learning}
	For unidirectional sequential recommendation, the task of predicting the next item tends to be adopted in their models. For example, these models create $N$-1 samples (like $([v_1], v_2)$ and $([v_1,v_2],v_3)$ from the original length $N$ behavioral sequence. But for the bidirectional sequential recommendation, if we also adopt this strategy to train model, these models create ($N$-1)! samples, which is time-consuming and infeasible. Thus we employ the mask task (same as \cite{ref_9,ref_11}) to efficiently train our model. Different from \cite{ref_9}, we add special tokens ``[UID]'' at the first of the sequence, which is used in the matching task. Here is a mask example in our model:
	\begin{eqnarray}
	&&\textbf{Input}: [\scriptstyle[{\rm UID}],v_1,v_2,v_3,v_4,v_5]\rightarrow [\scriptstyle[{\rm UID}],v_1,\scriptstyle{{\rm [mask]}}_2,v_3,\scriptstyle{[{\rm mask]}}_4,v_5], \nonumber\\
	&&\textbf{Labels}: \scriptstyle{[{\rm mask}]}_2=v_2,\scriptstyle{[{\rm mask}]}_4=v_4.\nonumber
	\end{eqnarray}
	where we randomly mask the proportion $\rho$ of all items in the input sequence (i.e., replace with special token ``[mask]''), and we always mask all of the successive same items at once to prevent the information leakage as far as possible. Our model needs to predict these masked items' ids based on their surrounding items. We define the negative log-likelihood loss for each masked input $S^{u'}$:
	\begin{equation}
	{Loss}_{mask}=\frac{1}{|S_{mask}^u|}\sum_{v_{mask}\in S_{mask}^u}-\log P_{mask}(v_{mask}=v_{mask}^*|S^{u'}),
	\label{eq6}
	\end{equation}
	where $S^{u'}$ is the masked version for user behavior history $S^u$, $S^u_{mask}$ is the random masked items in it, $v_{mask}^*$ is the label for the masked item $v_{mask}$, and the probability $P_{mask}(\bullet)$ is defined in Equation (\ref{eq3}). In multiple epochs, we produce different masked samples to train a more powerful model.
	
	Simultaneously, we add a new matching task into our model to capture user's global preference. We adopt the binary cross-entropy loss for each user: 
	\begin{equation}
	Loss_{matching}=-(\log(\sigma(score_{pos}\bullet c))+\log(1-\sigma(score_{neg}\bullet c)),
	\label{eq7}
	\end{equation}
	where $score_{pos}$ and $score_{neg}$ are defined in Equation (\ref{eq4}) and (\ref{eq5}) separately, $c$ is a scaling coefficient, which is assigned to 10 by us. In multiple epochs, we randomly generate $n$ negative items for each user sequence. And the total loss is the sum of the mask loss and the matching loss, shown as the following equation:
	\begin{equation}
	Loss=Loss_{mask}+Loss_{matching}.
	\label{eq8}
	\end{equation}
	
	We propose a new sample production method including three types. Firstly, we create samples used in the computation of the total loss. To address the mismatch between training and prediction, we create another type of sample that only masks the last item in the input user behavior sequences (same as \cite{ref_15}). To enhance our model's power of representations and alleviate the effect of mask noise, we also produce samples that are made of original sequences and the matching part. Here, we mix up these samples to train. Three types of samples are listed as follows:
	\begin{eqnarray}
	&&\textbf{Mask+Matching}: [[\scriptstyle[{\rm UID} ],v_1,\scriptstyle[{\rm mask}]_2,v_3,\scriptstyle[{\rm mask}]_4,v_5],[v_{pos},v_{neg_1},v_{neg_2}]],\nonumber\\
	&&\textbf{The last Mask}: [[\scriptstyle[{\rm UID}],v_1,v_2,v_3,v_4,v_5,\scriptstyle[{\rm mask}]_6],[]],\nonumber\\
	&&\textbf{Matching}: [[\scriptstyle[{\rm UID}],v_1,v_2,v_3,v_4,v_5],[v_{pos},v_{neg_1},v_{neg_2}]].\nonumber
	\end{eqnarray}
	
	In the prediction stage, we adopt a conventional strategy: sequential prediction (i.e., predicting the last item based on the final hidden representation of the sequence).
	
	\section{Experiments and Discussions}
	\subsection{Datasets and Baselines}
	We evaluate the proposed model on four representative datasets from three real-world applications, which vary significantly in domains and sparsity. \textbf{Amazon} \cite{ref_21} datasets contain product reviews and metadata from Amazon online shopping platform. They are separated into 24 categories according to the top level. In this work, we employ the small review subsets of ``Beauty'' and ``Video Games'' category. \textbf{Steam} \cite{ref_8} datasets contain reviews from the Steam video game platform. \textbf{MovieLens} \cite{ref_22} is a popular benchmark dataset, including several million movie ratings, reviews, etc. We employ the ``MovieLens-1M'' version.
	
	For the preprocessing procedure, we use a common strategy from \cite{ref_6,ref_8,ref_9,ref_16}. For all datasets, we transfer all ratings or reviews to implicit feedbacks (i.e., representing as numeric 1). Then, we group the interaction records by users and arrange them into sequences ordered by timestamp. We leave out users and items with fewer than five feedbacks. The statistics of the processed datasets are shown in Table \ref{tab1}. It needs to emphasize that we employ review datasets of Amazon rather than rating datasets, which is different from \cite{ref_8,ref_9}.
	
	\begin{table*}
		\centering
		\caption{Statistics of processed datasets.}\label{tab1}
		\begin{tabular}{p{60pt}<{\centering}p{35pt}<{\centering}p{35pt}<{\centering}p{35pt}<{\centering}p{50pt}<{\centering}}
			\toprule
			Dataset	&  $\scriptstyle\#$users  \quad   &  $\scriptstyle\#$items  \quad   &  $\scriptstyle\#$actions  \quad   &  Avg. length  \quad  \\
			\midrule
			Beauty & 22363 & 12101 & 0.23M & 6.88\\
			Video Games & 24303 & 10672 & 0.26M & 7.54\\
			Steam & 334730 & 13047 & 5.3M & 10.59\\
			ML-1M & 6040 & 3416 & 1.0M & 163.5\\
			\bottomrule
		\end{tabular}
	\end{table*}
	
	To verify the effectiveness of our method, we choose the following baselines: \textbf{POP} is a simple baseline that ranks items according to their popularity. \textbf{NCF} \cite{ref_16} is a general framework that replaces an inner product with a neural network to learn the matching function.
	\textbf{FPMC} \cite{ref_3} combines an MF term with first-order MCs to capture long-term preferences and short-term transitions respectively.
	\textbf{GRU4Rec+} \cite{ref_6} uses GRU with a new cross-entropy loss functions and sampling strategy to achieve session-based recommendation.
	\textbf{Caser} \cite{ref_19} employs CNN in both horizontal and vertical ways to model high-order MCs for the sequential recommendation.
	\textbf{SASRec} \cite{ref_8} uses a left-to-right Transformer language model to capture users' sequential behaviors.
	\textbf{BERT4Rec} \cite{ref_9} uses a two-layer Transformer decoder with the help of the Cloze task to mine bidirectional sequential information.
	
	For GRU4Rec+, Caser, SASRec, we use codes provided by the corresponding authors. For NCF, FPMC, and BERT4Rec, our model, we implement them by using $TensorFlow$. All parameters are initialized by using truncated normal distribution in the range [-0.02, 0.02]. We consider the $\ell 2$ weight decay from $\{1, 0.1, 0.01, 0.001\}$, and dropout rate from $\{0, 0.1, 0.2, ..., 0.9\}$, learning rate from $\{1e-3, 1e-4, 1e-5\}$, $\beta 1$ = 0.9, $\beta 2$ = 0.999, the hidden dimension size $d$ from $\{16, 32, 64\}$. For our model, we set the layer number $L = 2$ and head number $h = 2$ and set the maximum sequence length $N = 200$ for ML-1m, $N = 50$ for Beauty, Video Games, and Steam datasets, and we employ the same mask proportion $\rho$ with \cite{ref_9} (i.e., $\rho$ = 0.6 for Beauty and Video Games, $\rho$ = 0.4 for Steam, $\rho$ = 0.2 for MovieLens-1M). We consider the number negative sampling $n$ from $\{5, 10, 20\}$. All hyper-parameters are tuned on the validation sets. All results are under their optimal hyper-parameter settings.
	
	\subsection{Evaluation Metrics}
	To evaluate the performances of the recommendation models, we adopt the leave-one-out evaluation (i.e., next item recommendation) task, which is widely used in \cite{ref_6,ref_8,ref_9,ref_19}. For each user, we select the most recent action of his/her behavioral sequence as the test set, treat the second most recent action as the validation set, and utilize the remainder as the train set. Note that during testing, the input sequence is a combination of the train set and the validation set.
	
	We adopt Normalized Discounted Cumulative Gain (NDCG), Hit Ratio (HR) and Mean Reciprocal Rank (MRR) metrics as evaluation metrics. In this work, we report HR and NDCG with k =10. The higher value means better performance for all metrics. To avoid computing heavily on all item predictions, we randomly sample 100 negative items and rank these negative items with the ground-truth item for each user. Based on the rankings of these 101 items, the evaluation metrics can be evaluated.
	
	\subsection{Recommendation Performance}
	\begin{table*}
		\centering
		\caption{Recommendation performance. The best performing method in each row is boldfaced, and the second-best method in each row is underlined.}\label{tab2}
		\begin{tabular}{p{30pt}<{\centering}p{28pt}<{\centering}p{28pt}<{\centering}p{28pt}<{\centering}p{28pt}<{\centering}p{28pt}<{\centering}p{28pt}<{\centering}p{28pt}<{\centering}p{28pt}<{\centering}p{28pt}<{\centering}p{28pt}<{\centering}}
			\toprule
			Da- tasets & Metric & POP & NCF & FPMC & GRU4 Rec+ & Caser & SAS  Rec & Bert4 Rec & Ours & Improv. \\
			\midrule
			\multirow{3}{*}{Beaut} & NDCG $@$10 & 0.1793 & 0.2567 & 0.2937 & 0.2354 & 0.2705 & 0.3019 & \underline{0.3298} & \textbf{0.3370} & 2.1$\%$  \\
			y & HR$@$10 & 0.3363 & 0.4217 & 0.4064 & 0.3943 & 0.4223 & 0.4654  & \underline{0.4943} & \textbf{0.5009} & 1.3$\%$ \\
			& MRR & 0.1553 & 0.2229 & 0.2773 & 0.1105 & 0.2424 & 0.1865 & \underline{0.2431} & \textbf{0.2957} & 21.6$\%$ \\
			\midrule
			
			\multirow{3}{*}{Video} & NDCG $@$10 & 0.2512 & 0.3778 & 0.3225 & 0.4634 & 0.4137 & 0.4738 & \underline{0.4947} & \textbf{0.5163} & 4.1$\%$ \\
			Games & HR$@$10 & 0.4385 & 0.6031 & 0.5211 & 0.7137 & 0.6307 & \textbf{0.7320} & 0.6861 & \underline{0.7217} & -1.4$\%$ \\
			& MRR & 0.2151 & 0.3241 & 	0.2793 & 	0.2379 & 	0.3616 & 	0.3134 & 	\underline{0.4119} & 	\textbf{0.4396} & 	6.7$\%$ \\
			\midrule
			
			\multirow{3}{*}{Steam} & NDCG $@$10 & 0.4927 & 	0.4996 & 	0.5768 & 	0.5465 & 	0.5950 & 	0.6171 & 	\underline{0.6316} & 	\textbf{0.6460}	 & 2.2$\%$ \\
			& HR@10 & 	0.7556 & 	0.7629 & 	0.8216 & 	0.7986 & 	0.8310 & 	0.8440 & 	\underline{0.8633} & 	\textbf{0.8760} & 	1.4$\%$ \\
			& MRR & 	0.4225 & 	0.4295 & 	0.5087 & 	0.5247 & 	0.5292 & 	0.4177 & 	\underline{0.5488} & 	\textbf{0.5835} & 	6.3$\%$ \\
			\midrule
			
			\multirow{3}{*}{ML-} & NDCG $@$10  &  0.2455 & 	0.4094 & 	0.5258 & 	0.5456 & 	0.5408 & 	0.5354 & 	\underline{0.5483} & 	\textbf{0.5669} & 	3.3$\%$ \\
			1M & HR@10 & 	0.4458 &  0.6856 & 	0.7439 & 	0.7514 & 	0.7769 & 	\underline{0.7889} & 	0.7546 & 	\textbf{0.8096} & 	2.6$\%$ \\
			& MRR & 	0.2070 & 	0.3398 & 	0.3600 & 	0.4039 & 	0.4517 & 	0.3039 & 	\underline{0.4728} & 	\textbf{0.4973} & 	5.1$\%$ \\
			\bottomrule
		\end{tabular}
	\end{table*}

	Table \ref{tab2} illustrates the results of all methods on the four datasets. The last column is the improvements of our method relative to the best baseline. In our re-implementation of BERT4Rec, we reported different results compared to the original paper. The following three reasons need to be considered: firstly, we employ different datasets; secondly, we adopt a uniform negative sampling method instead of sampling according to items' popularity during the evaluation; thirdly, we reproduce it according to the published paper without the configuration file. From the results, we can summarize that:
	
	The non-personalized POP method gets the worst performance on all datasets because of just considering the number of interactions. In general, the sequential methods outperform traditional non-sequential methods such as NCF due to successive sequential information. This observation explains sequential information is beneficial to the improvement of recommendation performance. Particularly, on sparse dataset Video Games, FPMC performs worse than NCF only based collaborative filtering. This means that these datasets only have little additional sequential information and the neural model having more parameters is magnificent to recommendation performance. Among sequential recommendation baselines, on dense dataset ML-1m, Caser gets better performance than FPMC, which suggests that high-order interactions are useful for long input sequences. Furthermore, SASRec outperforms RNN (GRU4Rec+), CNN (Caser) sequential model, on the whole, meaning that the self-attention mechanism is more powerful for sequential feature extraction. BERT4Rec basically gets better performance than SASRec, suggesting that bidirectional sequential information is beneficial for the recommendation system. 
	
	According to the results, our method improved the best baseline on all four datasets w.r.t. the three metrics, especially, gaining 9.93$\%$ MRR improvements (on average) against the strongest baselines. Compared with BERT4Rec, an additional matching task and more abundant samples make our model outperform by a large margin w.r.t. the three metrics, which means the matching task is an important auxiliary tool to improve recommendation performance. 
	
\begin{figure}
		\centering
		\includegraphics[width=0.3\textwidth]{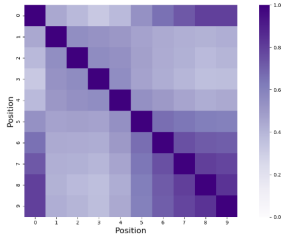}
		\caption{Heat-map of average correlation coefficients of output sequences on Beauty at different positions. The first position ``0'' denotes ``[UID]''.} \label{fig3}
	\end{figure}

	Meanwhile, Fig. \ref{fig3} visualizes average correlation coefficients of output sequences on Beauty of the first 10 items to qualitatively reveal the model's behavior. From the result, some tendencies can be concluded as follows: a) The users' representations are more affected by recent behavior, which is consistent with our common sense. Because recent items usually play a more important role in predicting the future. b) Items in our model tend to highlight the items on both sides, especially the surrounding items. This indicates bidirectional information has been mined successfully. 
	
	\subsection{Ablation Study}
	Finally, we use the ablation study to analyze numerous key components of our model to better understand their impacts. Table 3 shows the results of our default version and its variants on all four datasets (with d = 32).
	
	\begin{table*}
		\centering
		\caption{Ablation analysis (MRR). The bold score indicates performance better than the default version, while ``$\downarrow$'' indicates performance drop more than 10$\%$.}\label{tab3}
		\begin{tabular}{p{90pt}<{\centering}p{40pt}p{40pt}p{40pt}p{40pt}}
			\toprule
			Architecture & 	Beauty & 	V-Games & 	Steam & 	ML-1M \\
			\midrule
			Default & 	0.2957 & 	0.4396 & 	0.5835 & 	0.4973 \\
			w/o PE & 	0.2777 & 	0.3911$\downarrow$	 & 0.5591 & 	0.2955$\downarrow$ \\
			w/o Matching Task & 	0.2406$\downarrow$	 & 0.4101 & 	0.5462 & 	0.4513 \\
			1 head ($h = 1$) & 	0.2818 & 	0.4078 & 	0.5763 & 	0.4611 \\
			4 heads ($h = 4$)	 & 0.2826 & 	0.4216 & 	\textbf{0.5841} & 	\textbf{0.5012} \\
			1 layer ($L = 1$)	 & 0.2756 & 	0.4273 & 	0.5713 & 	0.4656 \\
			3 layers ($L = 3$) & 	\textbf{0.2981} & 	\textbf{0.4435} & 	\textbf{0.5907} & 	\textbf{0.5076} \\
			\bottomrule
		\end{tabular}
	\end{table*}
	
	\textbf{w/o PE}. Without the position embedding, the sequential model becomes the sequential model based on isolated actions. The attention weight on each item depends only on item embedding, which leads to the rapid decline of recommendation performance, especially on dense datasets, because long sequences have more noise actions.
	
	\textbf{w/o Matching Task}. The variant only adopts the mask task as an objective task (like BERT4Rec \cite{ref_9}). The recommendation performance witnesses a noticeable decrease on sparse datasets (e.g., Beauty) because the phenomenon of the mask dilemma is more common for short sequences. (i.e., more masked items mean less available context information and vice versa.)
	
	\textbf{Head number $h$}. Multi-headed attention can expand the model's ability to focus on different positions. We observe that long sequence datasets benefit from a larger $h$ (e.g., ML-1M), which means users' multiple interests are mined. 
	
	\textbf{Layer number $L$}. The results demonstrate that hierarchical Transformer layers can help model learn more complicated item transition patterns. This confirms the validity of the self-attention mechanism.
	
	\subsection{Space and Time Complexity Analysis}
	A theoretical analysis of the time and space complexity is presented as follows: 
	
	\textbf{Space complexity}. The learned parameters in our model are from the embedding layer, the transformer layers and the output layer. The total number of parameters is $O(|V|d+Nd+d^2)$, where $|V|$ means the number of the item set, $d$ is the size of hidden dimension, $N$ means the maximum sequence length. BERT4Rec \cite{ref_9} also has $O(|V|d+N d+d^2)$. 
	
	\textbf{Time complexity}. The time complexity of our model is mainly due to transformer layers, which is $O(dN^2+Nd^2)$. BERT4Rec has $O(|V|d+Nd+d^2)$. With GPU acceleration and $N=6$ in experiments, the difference is minor.
	
	\section{Conclusion}
	Recently, deep bidirectional sequential architecture proposed for neural language processing has brought impressive progress in recommender systems. In this paper, we optimize the bidirectional encoder representation recommender system via the additive matching task by the special token ``[UID]'' representing users. This method explicitly provides representations of users and captures user's global preference and main attentions in the sequence. Extensive experimental results on four real-world datasets indicate that our model outperforms state-of-the-art baselines. In the future, we will try to fuse heterogeneous interactions (e.g., purchase, review, clicks, etc.) in our model to achieve better performance.
	
	\section*{Acknowledgements}
	This work was partial supported by National Natural Science Foundation of China (Grant No. 41876098)
	
	%
	%
	%
	%
	
\end{document}